\documentclass[12pt]{article} 

\begin{document} 
\topmargin 0pt 
\oddsidemargin 0mm
\renewcommand{\thefootnote}{\fnsymbol{footnote}}
\begin{titlepage}
%\begin{flushright}
%\end{flushright}                                 
\vspace{5mm}

\begin{center}
{\Large \bf  Gravitational Thomas Precession - A Gravitomagnetic Effect ?} 
\vspace{6mm}

{\large Harihar Behera\footnote{email: harihar@iopb.res.in} }\\
\vspace{5mm}
{\em
 Department of Physics, Utkal University, Vanivihar, Bhubaneswar-751004, Orissa, India\\}
\vspace{3mm}
\end{center}
\vspace{5mm}
\centerline{{\bf {Abstract}}}
\vspace{5mm}
Gravitational Thomas Precession ( GTP ) is the name given to the Thomas
Precession when the acceleration is caused by a gravitational force
field. In continuation of our discussion on the idea of a GTP, in this 
note by way of considering the motion of a planet around the Sun, the GTP
is shown to be a gravitomagnetic effect that a planet might
experience  while  moving through a gravitational field (say that of
the Sun). The contribution of the  GTP to the perihelion advance of
Mercury is again estimated at  21.49 arc-seconds per century
confirming and clarifying our earlier results. \\
%\end{abstract}


\end{titlepage}
\section{Introduction}
The Thomas precession\cite{1,2,3} is purely kinematical in origin
\cite{2}. If a component of acceleration $ (\vec a)$ exists
perpendicular to the velocity $ \vec v $, for whatever reason, then
there is a Thomas Precession, independent of other effects
\cite{2}. When the acceleration is caused by a gravitational force
field, the corresponding Thomas Precession is reasonably referred to
as the Gravitational Thomas Precession (GTP). The question ``Is there
a gravitational Thomas Precession ?" was raised in \cite{4}  without
an answer. However, once the physics involved in the Thomas Precession is accepted, the possibility of the existence of  the GTP
in planetary motion can not be ruled out in principle. The GTP was
recently discussed in \cite{5}  wherein by introducing the GTP together
with the notion of  gravitational Lienard-Wiechart potentials (i.e., gravitational analogues of the electromagnetic Lienard-Wiechart potentials),
 the so called non-Newtonian perihelion shift of a planet (say Mercury)
 is seemingly explained, from a new angle, as a relativistic effect in
 flat-space  time. Surprisingly the estimated value for the perihelion
 advance in this new approach happens to coincide with Einstein's
 expression for the same effect, when a  spin-orbit factor 
 contributing insignificantly towards the observed effect, is
 reasonably dropped from the new expression. In the new approach the
 contribution of the GTP towards the perihelion advance came out as half of the
 Einstein's predicted value and the other half seemingly originate
 from what we call the GLWP( gravitational Lienard-Wiechart
 potentials). The aim of this note is to clarify and confirm our
 earlier results by illustrating  the  GTP effect as nothing but a gravitomagnetic effect. To this end a hypothetical situation involving the motion 
 of a comparatively low  mass point particle (call it a planet) around a
 massive central object (call it the Sun) might be considered as under.\\

\section{ The GTP as a Gravitomagnetic effect }     

Let the massive Sun with its rest mass $\,M_{\odot}\,$ be presumed
fixed at the origin of some frame S, and let a relatively light planet 
with rest mass $\,m_{0}\,$ move under the force of gravitational
interaction between them. In the frame S, the motion of the planet
(treated as a point particle) is described by its position as a
function of time $\,\vec r(t)\,$. At each instant $\,t\,$ the planet
will have a velocity $\,\vec v(t)\,=\,d\vec r/dt\,$, an acceleration
$\,\vec a(t)\,=\,d\vec v/dt\,$ as well as higher derivatives of
position, $\,\vec b(t)\,=\,d\vec a/dt\,$, etc. Over a very small time
interval  $\,t_{1}\,<\,t\,<\,t_{2}\,$, the planet is thought to be
moving under uniform acceleration. Then following Rohrlich \cite{6}
and Hill \cite{7}, the condition of uniform acceleration can be stated 
as :
\begin{equation}
\vec b\,+\,3{\gamma}^{2}\,(\vec a\cdot\vec v\,)\cdot\vec a\,=\,0
\end{equation}
where
\begin{equation} 
\gamma\,=\,(\,1\,-\,v^{2}/c^{2}\,)^{-1/2}
\end{equation}
 The most general motion $\,\vec r(t)\,$ in S with uniform
 acceleration is given by the general solution of Eq. (1). To fond this
 solution we note first that Eq.(1) can be written as
\begin{equation}
d(\gamma^{3}\,\vec a)/dt\,=\,0
\end{equation}
and can therefore be integrated immediately, $\,\gamma\,\geq\,1\,$, to 
yield
\begin{equation}
\gamma^{3}\vec a\,=\,\vec g
\end{equation} 
where $\,\vec g\,$ is a constant vector ( independent of $t$ )over a
very small time interval.
 A simple calculation shows that Eq.(4) can be put into the following
 form
\begin{equation}
d(\gamma\,\vec v)/dt\,-\,\gamma(\gamma^{2}\,-1\,)\,\vec v\times\,(\vec 
v\,\times\,\vec a)/v^{2}\,=\,\vec g.
\end{equation} 
This equation can be written as an equation of motion of the planet,
using 
\begin{equation}
\vec P\,=\,m_{0}\gamma\,\vec v,\,\,\,\,\,\,\vec F_{g}\,=\,m_{0}\vec g
\end{equation}
where  
\begin{equation}
\vec g\,=\,-\,GM_{\odot}\vec r/r^{3}
\end{equation}
 provided   one introduces ``a new force''
\begin{equation}
\vec F_{Th}\,=\,-(\,\gamma\,+\,1\,)\vec P\,\times\,\omega_{Th}\,\,\,\,\,\,\,\,\,\,\,{\mbox{where}}
\end{equation}

\begin{equation}
\vec \omega_{Th}\,=\,-\,(\gamma\,-\,1)\,\frac{\vec v\,\times\vec a}{v^{2}}
\end{equation}
is the angular  velocity associated with the Thomas precession. The
force $\,\vec F_{Th}\,$ may be designated as the Thomas force. Thus
the equation of the planet under uniform acceleration is 
\begin{equation}
\vec F\,=\,\frac{d\vec P}{dt}\,=\,\vec F_{g}\,+\,\vec F_{Th}.
\end{equation}
From Eqs.(2), (4), (7),(8) and (9) it can be shown that
\begin{equation}
\vec F_{Th}\,=\,m_{0}\,\vec v\,\times\,(\vec v\,\times\,\vec g)/c^{2}.
\end{equation}
so that the relativistic equation of motion of the planet becomes
\begin{equation}
\vec F\,=\,\frac{d\vec P}{dt}\,=\,m_{0}\vec g \,+\,m_{0}\,\vec
v\,\times\,(\vec v\,\times\,\vec g)/c^{2}\,=\,m_{0}\vec g
\,+\,m_{0}\,\vec v\,\times\,\vec B_{g}
\end{equation}
where 
\begin{equation}
\vec B_{g}\,=\,(\vec v\,\times\,\vec g)/c^{2}\,=\,\frac{GM_{\odot}}{c^{2}r^{3}}\,(\vec r \times \vec v )
\end{equation}
is what we may call the gravitomagnetic induction field\cite{8} felt by the
planet moving around the Sun in analogy with electromagnetism. It is
to be noted that the origin of this gravitomagnetic field lies in the
Thomas Precession in a gravitational field as per our derivation
presented here. Now introducing the angular momentum of the planet as 
 $\,\vec L\,=\,m(\vec r\times\vec v)\,$, Eq.(13) can be rewritten as
\begin{equation}
\vec B_{g}\,=\,\frac{GM_{\odot}}{mc^{2}r^{3}}\,\vec L
\end{equation}
If the planet has an orbital gravitomagnetic moment\cite{8} defined
by 
\begin{equation}
\vec {\mu}_{g}\,=\,\left(\frac{m_{0}}{2m}\right)\vec L\,,
\end{equation}
 then this would interact with the gravitomagnetic field (14) and the
 corresponding interaction energy according to Maxwellian Gravity \cite{8}
would be
\begin{equation}
U_{gm}\,=\,-\,\vec {\mu}_{g}\,\cdot\,\vec B_{g}\,=\,-\frac{GM_{\odot}m_{0}}{2c^{2}r^{3}}\,\frac{L^{2}}{m^{2}}
\end{equation}
But it can be shown that $\,
\frac{L^{2}}{m^{2}}\,=\,\frac{{L_{0}}^{2}}{{m_{0}}^{2}}\,$, where
$\,\vec L_{0}\,=\,m_{0}(\,\vec r\,\times\,\vec v\,)$,( see the
Appendix-I ), so that Eq.(16) can be reduced to 
\begin{equation}
U_{gm}\,=\,-\,\vec {\mu}_{g}\,\cdot\,\vec B_{g}\,=\,-\frac{GM_{\odot}}{2m_{0}c^{2}r^{3}}\,{L_{0}}^{2}
\end{equation}
This is what represents the $ \,\vec L\,\cdot\,\vec\omega_{gT}\,$ term 
in Eq.(29) of \cite{5}, where  $\vec\omega_{gT}$ represents the
gravitational Thomas Precession frequency corresponding to Eq.(24) of
\cite{5} with the non-relativistic assumption that $ \vec
L\,\simeq\,\vec L_{0}\,$. Then the rest of the arguments regarding
the Thomas Precession's contribution to the perihelion advance follows
as given in \cite{5}. The result is that this contribution amounts to 
$\,21.49\,$ arc-seconds per century for Mercury. This seems to be a
clarified version of our earlier results in \cite{5}.The above
discussion seems to look at the Gravitational Thomas Precession as a
Gravitomagnetic effect in flat space-time.

\textbf{Acknowledgments}\\   
 The author is very much grateful to Prof. N. Barik and
 Prof. L. P. Singh, both of Utkal University, Vanivihar, Bhubaneswar
 and Dr. P. C. Naik, D. D. College, Keonjhar for their valuable
 suggestions for the improvement of this letter. The author also
 acknowledges  the help received from Institute of Physics,
 Bhubaneswar for using its Library and Computer Centre for this
 work.\\
 {\bf{Appendix-I}}

Let us start with the force Eq.(12),viz.,
\begin{equation}
\vec F\,=\,\frac{d\vec P}{dt}\,=\,m_{0}\vec g \,+\,m_{0}\,\vec
v\,\times\,(\vec v\,\times\,\vec g)/c^{2}
\end{equation}
This equation can be reduced to 
\begin{equation}
\vec F\,=\,\frac{d\vec P}{dt}\,=\,m_{0}\vec g\,(\,1\,-\,v^{2}/c^{2}\,) \,+\,m_{0}\,\vec v\,(\vec v\,\cdot\,\vec g)/c^{2}
\end{equation}
Thus in relativistic central force Kepler problem we find the force is not 
strictly central , since it is evident from Eq.(19) that a component
of the force exists in the direction of the velocity vector $\vec v
$. Therefore the torque acting on the planet in the central force
Kepler problem would be given by
\begin{equation}
\vec N\,=\,\frac {d\vec L}{dt}\,=\,\vec r\,\times\,\vec F\,=\,m_{0}(\vec v\,\cdot\vec g)(\,\vec r\,\times\,\vec v\,)/c^{2}\,\neq\,0
\end{equation}
for non-radial motion. With $\,\vec L\,=\,m(\,\vec r\,\times\,\vec
v)\,$ and knowing that 
\begin{equation} 
c^{2}\frac {dm}{dt}\,=\,m_{0}(\vec v\,\cdot\vec g)
\end{equation}
Eq.(20) can be written as 
\begin{equation}
\frac {d\vec L}{dt}\,=\,\frac {dm}{dt}\frac{\vec L}{m}
\end{equation}
from which we can get 
\begin{equation}
\int_{L_{0}}^{L}\frac {dL}{L}\,=\,\int_{m_{0}}^{m}\frac {dm}{m}.
\end{equation}
This equation on integration yields
\begin{equation}
\ln{\frac{L}{L_{0}}}\,=\,\ln{\frac{m}{m_{0}}}
\end{equation}
implying that $ \frac{L}{L_{0}}\,=\,\frac{m}{m_{0}} $,    or
\begin{equation}
\frac {L}{m}\,=\,\frac {L_{0}}{m_{0}}\,= constant
\end{equation}
which is our required equation.

%\textbf{References} 


\begin{thebibliography}{99} 
\bibitem{1} L. T. Thomas, Phil. Mag., 3, 1 (1927). 
\bibitem{2} J. D. Jackson,\textit{ Classical Electrodynamics,}
 2nd Ed. (Wiley, New York, 1975).  
\bibitem{3} H. Goldstein,\textit{ Classical Mechanics,} 2nd Ed.( Narosa
  Publishing House, New Delhi, 1995 ).    
\bibitem{4} Robert T. Jantzen, Paolo Carini, and  Donato Bini,
 `` Gravitomagnetism : Relativity of Splitting Formalisms'',
 Procs. Sixth Marcel Grossmann Meeting on General Relativity(1991), Eds
 H. Sato and T. Nakamura,( World Scientific, Singapore,1993),
 pp.135-137.
\bibitem{5} H. Behera and P. C. Naik, `` A flat space-time
  relativistic explanation for the perihelion advance of
  Mercury'', 
[ See the LANL E-print : astro-ph/0306611].
\bibitem{6} F. Rohrlich, Ann. of Phys., 22, 169-191 (1963).
\bibitem{7} E. L. Hill, Phy. Rev. 72, (1947) 143.
\bibitem{8} H. Behera and P. C. Naik, ``Gravitomagnetic Moments and
  Dynamics of Dirac ( spin 1/2 ) fermions in flat space-time
  Maxwellian Gravity'', To appear in Int. J. of Mod. Phys. A. 
[ see the LANL E-print : gr-qc/0304084 v2 ].
\end{thebibliography}
\end{document}